\documentclass{article}
\usepackage{spconf,amsmath,graphicx, cite}

\newcommand{\tabref}[1]{\mbox{Table~\ref{#1}}}
\newcommand{\figref}[1]{\mbox{Figure~\ref{#1}}}
\newcommand{\secref}[1]{\mbox{Section~\ref{#1}}}

\usepackage{amsfonts}
\usepackage{url}
\usepackage{color}
\usepackage[english]{babel}
\usepackage[T1]{fontenc}
\usepackage[utf8]{inputenc}
\usepackage{booktabs} 
\usepackage{algorithm}
\usepackage{algorithmic} 
\usepackage{nicefrac}

\newcommand{\etal}{\textit{et al.}}
\newcommand{\ourname}{ByteCover}

\title{ByteCover: Cover Song Identification via Multi-Loss Training}
\name{Xingjian Du$^{1}$, Zhesong Yu$^{1}$, Bilei Zhu$^{1}$, Xiaoou Chen$^{2}$, Zejun Ma$^{1}$}
\address{$^{1}$ Bytedance AI Lab \qquad $^{2}$ Peking University} 
\begin{document}
%
\maketitle
\begin{abstract}
We present in this paper ByteCover, which is a new feature learning method for cover song identification (CSI). ByteCover is built based on the classical ResNet model, and two major improvements are designed to further enhance the capability of the model for CSI. In the first improvement, we introduce the integration of instance normalization (IN) and batch normalization (BN) to build IBN blocks, which are major components of our ResNet-IBN model. With the help of the IBN blocks, our CSI model can learn features that are invariant to the changes of musical attributes such as key, tempo, timbre and genre, while preserving the version information. In the second improvement, we employ the BNNeck method to allow a multi-loss training and encourage our method to jointly optimize a classification loss and a triplet loss, and by this means, the inter-class discrimination and intra-class compactness of cover songs, can be ensured at the same time. A set of experiments demonstrated the effectiveness and efficiency of ByteCover on multiple datasets, and in the  Da-TACOS dataset, ByteCover outperformed the best competitive system by 18.0\%.
\end{abstract}

\begin{figure*}
	\centering
	\includegraphics[width=0.96\textwidth]{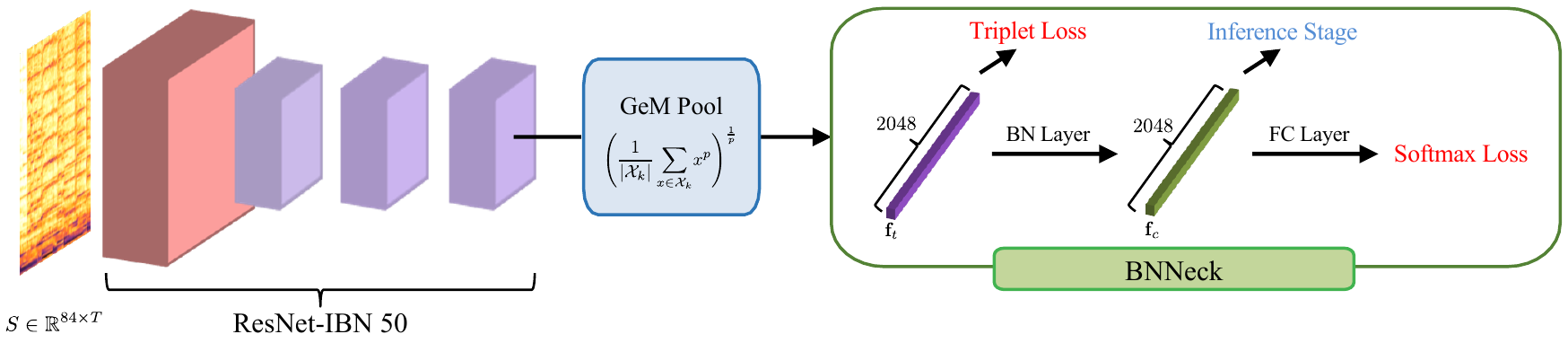}
	\caption{Model structure of ByteCover.}
	\label{fig:net}
\end{figure*}
\section{Introduction}
\label{sec:intro}
Cover song identification (CSI), which aims at finding cover versions of a given music track in a music database, is an important task in the field of music information retrieval (MIR), with a number of potential applications such as music organization, retrieval, recommendation and music license management. However, cover versions may be different from the original music in many aspects such as key, tempo, structure, instrumentation or even genre, which makes CSI a challenging problem \cite{serra2010audio}.



Traditional CSI methods based on the extraction of hand-crafted features (e.g., predominant melody, chroma sequence, chord progression, and harmonic pitch class profiles (HPCP)) and the use time series matching techniques (e.g., dynamic time warping (DTW)) have achieved promising results on small datasets \cite{marolt2006mid, ellis2007identifying, serra2008chroma, serra2009cross}. However, most of these methods contain exhaustive iteration and matching, which makes it difficult for them to scale to large datasets consisting of hundreds of thousands of or even millions of music tracks \cite{xu2018effective, yesiler2020accurate}. This limits the use of CSI in real-world industry applications. Recently, with their popularity in computer vision and other MIR tasks, deep learning methods have been introduced to solve the CSI problem. Roughly speaking, existing deep learning methods to CSI can be classified into two categories. The first category of methods, e.g., \cite{xu2018key, yu2019temporal, yu2020learning}, treats CSI as a multi-class classification problem, where each version group is considered as an unique class. Convolutional neural networks (CNNs) are trained to classify music tracks in the training set and during retrieval, the network's penultimate layer is used to generate feature for audio matching. The second category of methods, on the other hand, treats CSI as a metric learning problem. These methods, e.g., \cite{doras2019cover, yesiler2020accurate}, mostly use triplet loss as the objective function and train CNN-based models to minimize the distances between cover pairs and maximize the distances between different covers. After training, the convolutional part of the network is used as feature extractor for the further identification.

Experiments in the existing works have proved that deep learning models can boost the performance of CSI by a large margin over traditional methods. Moreover, the feature representations learned by deep models are usually fixed-length vectors where the length is unrelated to the original music track. This raises an advantage that the similarity between features can be directly measured by Euclidean distance or Cosine distance, and therefore the features can be efficiently indexed and retrieved using existing nearest-neighbor search libraries, even for a large dataset. In this paper, following the success of deep learning models introduced above, we present the Bytedance Cover Song Identification System (ByteCover), which is a new feature learning system that aims at learning fixed-length feature representations for CSI. ByteCover is built based on the classical CNN model ResNet50 \cite{he2016deep}, and includes two major improvements that make the model more powerful for CSI. Together with other minor designs such as generalized mean pooling (GeMPool), these two improvements helped ByteCover beat all the existing state-of-the-art methods on all the widely-used datasets in our experiments. Especially, on the Da-TACOS dataset, ByteCover achieved a mean average precision of 71.4\%, which is 18.0\% higher than that the best competitive system.

Our first improvement is that we introduce instance normalization (IN) to the problem of CSI and combine IN with batch normalization (BN) to build the IBN block and further the ResNet-IBN model for CSI. The combination of IN and BN was first introduced in computer vision \cite{pan2018two} to joint utilize the ability of IN for learning features that are invariant to appearance changes, such as colors, styles, and virtuality/reality, and the ability of BN for preserving content related information. In CSI, we also have the demand of keeping the version information of music, while designing features that are robust to the changes of key, tempo, timbre, and genre, among others. We believe that the use of IBN would also be beneficial for CSI, and our ablation experiments will show that, by integrating IN with BN in ResNet50, the performance of CSI can be significantly improved.

Our second improvement is that, different from existing works which either formulate CSI as a classification problem or treat it as a metric learning problem, ByteCover attempts to learn features by jointly optimizing a classification loss and a triplet loss. It has been testified that classification loss emphasizes more on inter-class discrimination while triplet loss emphasizes more on intra-class compactness \cite{taha2020boosting}. In the literature, a few studies have been conducted to combine these two losses, and the experimental results showed that the combination can result in better performance than using a single loss, in tasks such as person re-identification \cite{luo2019bag}, fine-grained visual recognition and ego-motion
action recognition \cite{taha2020boosting}. In this paper, we investigate the use of BNNeck proposed in \cite{luo2019bag} to fuze the classification and triplet losses for CSI, and by using this multi-loss training, feature representations that are both robust and discriminative can be learned.

\section{ByteCover Approach}
\label{sec:approach}


Our aim of using ByteCover is to derive from each input music track a single global feature that encodes the version information, and by comparing the features of any two music tracks using Cosine distance, we can obtain an estimation of whether these two music tracks are cover versions of each other. \figref{fig:net} shows the model structure of ByteCover. As shown in the figure, ByteCover takes as input a constant-Q transform (CQT) spectrogram and employs a CNN-based model, i.e., ResNet-IBN, for feature learning. The learned feature map is then compressed by GeMPool to form a fixed-length vector, which is then used to calculate the triplet loss and Softmax loss jointly. The joint training of these two losses are achieved by using the BNNeck.


\par


\vspace{-1em}
\subsection{ResNet-IBN}



For the simplification of the overall pipeline, we use the CQT spectrogram rather than the sophisticated features widely used in the CSI \cite{yesiler2020accurate}. 
In terms of the parameters of the CQT, the number of bins per octave is set as $12$ and Hann window is used for extraction with a hop size of $512$. Besides this, the audio is resampled to 22050 Hz before the feature extraction. Afterwards, the CQT is downsampled with an averaging factor of 100 along the temporal dimension.  This size reduction of input feature improves the efficiency of the model and reduces the latency of our CSI system. As a result, the input audio is processed to a compressed CQT spectrogram $S \in \mathbb{R}^{84 \times T}$, where the $T$ depends on on the duration of the input music.

In the CSI task, the robustness against the change of various musical elements is highly related to the model's generalization on large-scale music corpus. At the same time, we also hope that this robustness will not undermine the ability of the model to identify covers.
Inspired by the success of instance batch normalization~(IBN) networks \cite{pan2018two} in learning an invariant representation of an image and preserving the discrimination for person re-identification, we replaced the residual block in the vanilla ResNet \cite{he2016deep} with IBN block.\par

\begin{figure}
	\centering
	\includegraphics[width=0.45\textwidth]{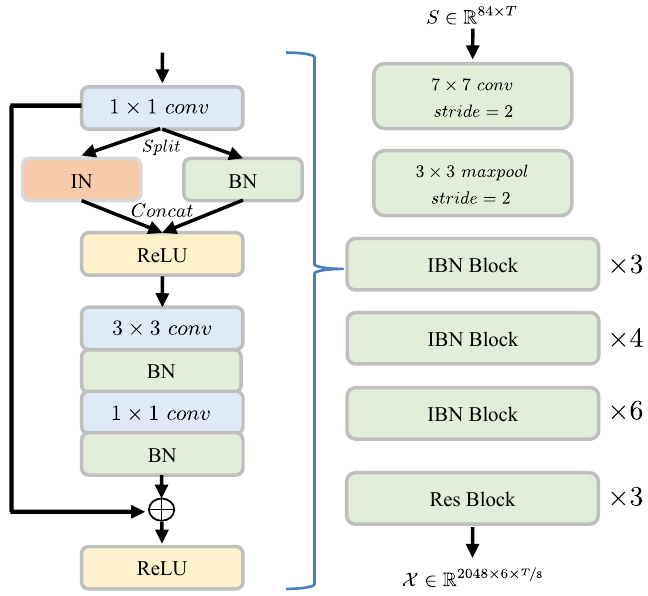}
	\caption{The Structure of IBN Block and ResNet-IBN 50.}

	\label{fig:ibn}
\end{figure}

The original ResNet \cite{he2016deep} consists of four stages, each having 3, 4, 6, and 3 basic residual blocks in turn. In addition, the first stage has a $7 \times 7$ convolution and $3 \times 3$ pooling layer to process the input. To transform a ResNet to a model equipped with IBN module for learning an invariant embedding, the residual block which is the basic elements of the model, are replaced with IBN blcok. As \figref{fig:ibn} shows, the BN is applied to half of the channels and IN normalized the other channels after the first convolution for every residual block.  The IN modules allow the model to learn style invariant features for music so that it could make better use of the music performances with a high diversity of styles within a single dataset. Besides, to prevent that the model's learning capacity is degraded by too many IN layers, the last module has not been modified and remains the same. Furthermore, to enlarge the spatial size of the output feature map, the stride of the first convolution layer in the last residual block is changed to $1$. Thus the input spectrogram has been processed to a 3-D embedding ${\cal X} \in \mathbb{R}^{K \times H \times W}$  after being downsampled four times and the channel expansion.  Here, $K=2048$ is the number of output channel, $H=6$ and $W=\nicefrac{T}{8}$ are the spatial size along the frequency and time axis, respectively. \par

To compress the feature map ${\cal X} $ to a fixed-length vector $\textbf{f}$, a spatial aggregation module is needed. Average pooling and max pooling are two common methods for aggregating local feature, and the several CSI models \cite{yu2019temporal,yu2020learning} attach a average pooling layer after the feature extractor. To boost the performance of our CSI model, we adopt a novel GemPool module, which unifies these two aggregation methods. The compression results $\textbf{f}$ can be given by 
\vspace{-0.6em}
\begin{equation}
\textbf{f} = [f_1 \ldots f_k \ldots f_K]^\top, f_k =\! \left( \frac{1}{|{\cal X}_{k}|}\sum_{x \in {\cal X}_{k}}\!x^{p} \right)^\frac{1}{p} \textit{.}
\label{equ:avgpooling}
\end{equation}
The behavior of the GeM pooling can be controlled by adjusting parameter $p$ : the GeM pooling is equivalent to the average pooling when $p = 1$ and max pooling for $p \rightarrow \infty$. The $p$ is set to a trainable parameter to fully exploit the advantage of deep learning.


\subsection{BNNeck for Multi-Loss Training}

In the field of CSI, there are two paradigms of training the model: they are classification and metric learning paradigms \cite{doras2020combining}. The former \cite{yu2020learning} uses an additional fully connected layer as a classifier to learn embeddings for cover detection through a proxy classification task. The latter \cite{yesiler2020accurate} directly uses metric learning methods such as triplet loss to make the Euclidean distance between cover pairs smaller than non-cover pairs.  

A straightforward method to merge these two training paradigms is calculating the classification loss and the triplet loss on the same feature.
However, the experimental results in \secref{sec:exp} show that this naive approach cannot effectively improve the performance of the CSI model.
The disharmony between classification loss and triplet loss may be the reason for the unsatisfactory performance of joint learning. The work of Hao \etal \cite {luo2019bag} indicates that the targets of these two losses are inconsistent in the embedding space. 

For sample pairs in the embedding space, classification loss mainly optimizes the cosine distance while triplet loss focuses on the Euclidean distance. If we use these two losses to optimize a feature vector simultaneously, their goals may be inconsistent. Consequently, triplet loss may influence the clear decision surfaces of classification loss, and classification loss may reduce the intra-class compactness of triplet loss \cite {luo2019bag}. 

Thus the BNNeck module inserts a BN layer before the classifier,  which is a no-biased FC layer with weights $\textbf{W}$, as \figref{fig:net} depicts. The feature yield by the GeM module is denoted as $\textbf{f}_t$ here. We let $\textbf{f}_t$ pass through a BN layer to generate the feature $\textbf{f}_c$. In the training stage, $\textbf{f}_t$ and $\textbf{f}_c$ are used to compute triplet and classification losses, respectively.  

Because the normalized features are more suitable for the calculation of cosine metric, we use the $\textbf{f}_c$ to do the computation of the similarity among the performances during the inference and retrieval phases. Overall, the objective function used for training our model can be derived by
\begin{align}
    L &= L_{cls}(\mathbf{f}_c) + L_{tri}(\textbf{f}_t) \\
       &= CE(Softmax(\mathbf{W}\textbf{f}_c), y) + [d_p - d_n + \alpha]_{+} ~\textit{,}
\end{align}
where $d_p$ and $d_n$ are feature distances of positive pair and negative pair.
$\alpha$ is the margin of triplet loss, and $[z]_{+}$ equals to $max(z,0)$.
In this paper, $\alpha$ is set to $0.3$. In terms of the classification loss $L_{cls}$, the $CE$ refers the cross entropy function and $y$ is the ground truth label. Finally, we train the model with Adam optimizer. The batch size and the learning rate are $32$ and $0.0004$, respectively.
\section{Experiments}
\label{sec:exp}

\begin{table}
  \small
  \centering
  \setlength\tabcolsep{12pt}
  \label{tab:com}
  \begin{tabular}{l c c c}
      \toprule
      & MAP & P@10 & MR1  \\ 
      \midrule
      &\multicolumn{3}{c}{Results on Youtube350} \\
      \hline
      SiMPle \cite{silva2016simple} & 0.591 & 0.140 & 7.91  \\ 
      TPPNet \cite{yu2019temporal}& 0.859 & 0.188 & 2.85  \\
      CQT-Net \cite{yu2020learning} & 0.917 & 0.192 & 2.50  \\
      MOVE~\cite{yesiler2020accurate}  & 0.888 & - & 3.00 \\ 
      \textbf{ByteCover}  & \textbf{0.955} & \textbf{0.196} & \textbf{1.60}  \\ 
      \hline
      &\multicolumn{3}{c}{Results on Covers80} \\
      \hline
      Qmax \cite{serra2009cross} & 0.544 & 0.061 & - \\
      TPPNet \cite{yu2019temporal}& 0.744 & 0.086 & 6.88  \\
      CQT-Net \cite{yu2020learning} & 0.840 & 0.091 & 3.85  \\ 
      \textbf{ByteCover} & \textbf{0.906} & \textbf{0.093} & \textbf{3.54}  \\ 
      \hline
      &\multicolumn{3}{c}{Results on Da-TACOS} \\
      \hline
      Qmax~\cite{serra2009cross}  & 0.365 & - & 113  \\
      SiMPle\cite{silva2016simple} & 0.332 & - & 142  \\
      MOVE \cite{yesiler2020accurate} & 0.507 & - & 40  \\
      Re-MOVE \cite{yesiler2020less} & 0.534 & - & 38  \\
      \textbf{ByteCover} & \textbf{0.714} & \textbf{0.801} & \textbf{23.0}  \\
      \hline
      &\multicolumn{3}{c}{Results on SHS100K-TEST} \\
      \hline
      Ki-CNN \cite{xu2018key}& 0.219 & 0.204 & 174  \\ 
      TPPNet \cite{yu2019temporal} & 0.465 & 0.357 & 72.2 \\
      CQT-Net \cite{yu2020learning} & 0.655 & 0.456 & 54.9  \\ 
      \textbf{ByteCover} & \textbf{0.836} & \textbf{0.534} & \textbf{47.3}  \\ 
      \bottomrule
  \end{tabular}
  \caption{Performance on different datasets (- indicates the results are not shown in original works).}
  \end{table}
\subsection{Evaluation Settings}
\label{subsec:dataset}

To evaluate the performance of our \ourname\ model, we conducted several experiments on four publicly available
benchmark datasets, including SHS100K \cite{xu2018key}, Youtube350 \cite{silva2015music}, Covers80 \cite{ellis2007identifying}, and Da-TACOS    \cite{yesiler2019datacos}. \par

\textit{SHS100K}, which is collected from \textit{Second Hand Songs website} by \cite{xu2018key}, consisting of 8858 songs with various covers and 108523 recordings. For this data, we follow the settings of \cite{yu2020learning}. The dataset is split into the training, validation, and test sets with a ratio of 8:1:1. Our \ourname\ model was trained on the training subset of SHS100K, and the results on the test subset are reported.\par

\textit{Youtube350} \cite{silva2015music} is collected from the YouTube, containing 50 compositions of multiple genres \cite{silva2015music}. Each song in Youtube350 has 7 versions, with 2 original versions and 5 different versions and thus results in 350 recordings in total. In our experiment, we used the 100 original versions as references and the others as queries following the same as \cite{silva2016simple,yu2019temporal,xu2018key}. \par

\textit{Covers80} \cite{ellis2007identifying} is a widely used benchmark dataset in the literature. It has 80 songs, with 2 covers for each song, and has 160 recordings in total. To compare with existing methods, we computed the similarity of any pair of recordings.

\textit{Da-TACOS} \cite{yesiler2019datacos} consists of 15000~performances, of which 13000~performances belong to 1000~cliques, with 13~samples in each clique; the remaining 2000~pieces do not belong to any clique.
It is worth mentioning that there are 8373 samples of the Da-TACOS which are also belongs to the training set of SHS100K. Thus we removed these samples from the training set and report the results of retrained model.


During the retrieval phase, the cosine distance metric was used to estimate the similarity between two musical performances.
Following the evaluation protocol of the Mirex Audio Cover Song Identification Contest \footnote{\url{https://www.music-ir.org/mirex/wiki/2020:Audio_Cover_Song_Identification}}, the mean average precision (mAP), precision at 10 (P@10), and the mean rank of the first correctly identified cover (MR1) are reported as evaluation metrics.


\vspace{-1em}
\subsection{Comparison and Ablation Study}
\label{ssec:comp}

\begin{table}[]
\small
\centering
\begin{tabular}{@{}ccccccc@{}}
\toprule    & 128 & 256 & 2048 & Move, 16000 & ReMove, 256  \\ \midrule
mAP &   0.615    & 0.653 & \textbf{0.714}  & 0.507         & 0.524                   \\ \bottomrule
\end{tabular}
\vspace{-0.6em}
\caption{mAP on Da-Tacos for different embedding sizes $K$.}
\label{tab:size}
\end{table}
\begin{table}[]
\small
\centering
\begin{tabular}{@{}lcc@{}}
\toprule
         & mAP on Da-TACOS & mAP on SHS100K-TEST \\ \midrule
Baseline & 0.546          & 0.683               \\
+~IBN     & 0.617             & 0.734               \\
+~BNNeck  & 0.698           & 0.802               \\
+~GeMPool & \textbf{0.714}           & \textbf{0.836}               \\ \bottomrule
\end{tabular}
\vspace{-0.8em}
\caption{The ablation study of ByteCover. Baseline is a ResNet based network trained by the classification loss only. The improvements are aggregated in turn.}
\vspace{-0.3em}
\label{tab:ablation}
\end{table}

We compared \ourname\ against published state-of-the-art methods on four datasets, which are mentioned above. 
Our \ourname\ model outperformed all previous methods, achieving a new state-of-the-art performance by 18.0\% mAP over Re-MOVE \cite{yesiler2020less} on the largest benchmark dataset Da-TACOS. We found that our model can often achieve more significant advantages on datasets with more samples, more varied styles, and more diverse genres, like SHS100K or Da-TACOS.  \par
To analyze the importance of different parts of the ByteCover system quantitatively, we trained the models with several settings. The baseline was the original ResNet-50 based model trained by the classification loss only, and a global average pooling module produced the final output. Other settings and results are presented in \tabref{tab:ablation}. 

To prove the scalability of \ourname\ model and a fairer comparison with Re-MOVE \cite{yesiler2020less}, we report our method's performances with the embedding vectors of different lengths. We utilized a simple fully-connected layer as the projection head to reduce feature vector size from 2048 to 128 or 256, rather than the sophisticated knowledge distillation used in \cite{yesiler2020less}. As the \tabref{tab:size} shows, our model still surpasses the state-of-the-art method Re-MOVE~\cite{yesiler2020less} by a large margin while the length of our feature vector is half of \cite{yesiler2020less}.

\section{Conclusion}
In this paper, we propose a simple and efficient design of the cover song identification system.
The combination of classification and metric learning training schemes encourages our model to learn a discriminative feature embedding. The utilization of IBN leads the model to be robust against the musical variations. The results show that our \ourname\ system outperforms all previous state-of-the-art cover detection methods by a significant margin on four public benchmark datasets.   


\vfill\pagebreak

\bibliographystyle{IEEEbib}
\bibliography{6_refs}

\end{document}